\documentclass [
 aip,
 jmp,%
 amsmath,amssymb,
 reprint,%
]{revtex4-1}

\usepackage{graphicx}
\usepackage{dcolumn}
\usepackage{bm}
\usepackage{esvect}
\usepackage{lipsum}
\usepackage{float}
\usepackage{color}

\begin{document}
\preprint{AIP/123-QED}

\title[Long-Lived Non-Equilibrium Interstitial-Solid-Solutions in Binary Mixtures]{Long-Lived Non-Equilibrium Interstitial-Solid-Solutions in Binary Mixtures}

\author{Ioatzin R\'ios de Anda}
\affiliation{H.H. Wills Physics Laboratory, Tyndall Ave., Bristol, BS8 1TL, UK}
\author{Francesco Turci}
\affiliation{H.H. Wills Physics Laboratory, Tyndall Ave., Bristol, BS8 1TL, UK}
\author{Richard P. Sear}
\affiliation{Department of Physics, University of Surrey, Guildford, Surrey GU2 7XH, UK}
\author{C. Patrick Royall}
\affiliation{H.H. Wills Physics Laboratory, Tyndall Ave., Bristol, BS8 1TL, UK}
\affiliation{School of Chemistry, Cantock's Close, University of Bristol, BS8 1TS, UK}
\affiliation{Centre for Nanoscience and Quantum Information, Tyndall Avenue, Bristol BS8 1FD, UK}

\date{\today}

\begin{abstract}
We perform particle resolved experimental studies on the heterogeneous  crystallisation process of two component mixtures of hard spheres. The components have a size ratio of 0.39.
We compared these with molecular dynamics simulations of homogenous nucleation.
We find for both experiments and simulations that the final assemblies are 
interstitial solid solutions, where the large particles form crystalline close-packed lattices, whereas the small particles occupy random interstitial sites.
This interstitial solution resembles that found at equilibrium when the size ratios are 0.3 [Filion \textit{et al.}, Phys. Rev. Lett. \textbf{107}, 168302 (2011)] and 0.4 [Filion, PhD Thesis, Utrecht University (2011)].
However, unlike 
these previous studies, for our system simulations showed that the small particles are trapped
in the octahedral holes of the ordered structure formed by the large particles, leading to long-lived non-equilibrium structures in the time scales studied and not the equilibrium interstitial solutions found
earlier. Interestingly, the percentage of small particles in the crystal formed by the large ones rapidly reaches a maximum of $\sim$14\% for most of the packing fractions tested, unlike previous predictions where the occupancy of the interstitial sites increases with the system concentration. Finally, no further hopping of the small particles was observed.
\end{abstract}

\maketitle


\section{\label{sec:level1}Introduction}

Solid solutions are some of the toughest and
most versatile crystalline materials \cite{cahn2009}.
Steel is a hugely solid solution
\cite{verhoven07book}, it is an interstitial solid solution (ISS)
in which the iron atoms form an ordered crystal lattice, with the smaller
carbon atoms in the interstices.
 Substitutional solid solutions (SSS)
are also important, there the second component is not in the interstices
between atoms of the first component, but substitutes for atoms of the first
component, on the lattice.
Very recent developments in electron microscopy
have for the first time allowed us to analyse the disorder of an SSS of
iron in platinum \cite{yang17}. Also, High Entropy Alloys (HEAs),
which are effectively a multicomponent SSS, are currently very actively
studied as they are among the toughest materials known \cite{gludovatz14,Filion2011,Filion2011b,Zhang2014,Royall2015}.

Hard spheres, as epitomised by colloids,
are widely used as 
models to study the  self-assembly and phase behaviour processes of atoms and molecules. Since the structural evolution of colloidal suspensions can be followed in real space by means of confocal imaging, their study has allowed us to monitor the crystallisation process of different systems\cite{Ivlev2012}. Additionally, the equilibrium structures of such processes have also been successfully characterised in simulations, which is crucial for fundamental understanding and for many applications, including material science, metallurgy, and biotechnology\cite{Sandomirski2011, Saunders2005}. The most popular example of these systems is monodisperse hard spheres, where the particles interact only by hard-core repulsions and whose phase behaviour has been widely studied, through both simulations and experiments\cite{Filion2011b}. These studies have identified the face-centered cubic (fcc) -often in a random mixture with hexagonal close packing (hcp)-, as the solid stable structure\cite{Schofield2005,Filion2011}.

Crystallisation of multicomponent colloidal mixtures has acquired more interest, as these present a richer and more complex phase behaviour than their monocomponent counterpart \cite{Pham2016, Schofield2005, Trizac1997, Filion2011, Hunt2000,Bartlett1992,Leocmach2010}. Amongst these systems, we find binary mixtures of hard spheres, which are composed of two populations of spheres of differing sizes. Similar to one-component hard spheres, they are also the simplest multicomponent colloidal system and thus, form a reference model to study phase transitions of mixtures\cite{Schofield2005,Trizac1997,Filion2011,Hunt2000}. 
Crystals of these systems were 
observed by Sanders and Murray back in 1978, whilst analysing the microscopic structure of gem opals, which are composed of silica colloids and had an structure analogue to AlB$_{2}$ and NaZn$_{13}$ \cite{Sanders1978}. So far, both experimental and simulation studies have identified that the phase diversity found in binary mixtures depends on the size ratio of the components, the number ratio and total concentration of the particles \cite{Filion2011b, Filion2011, Vermolen2009, Christova2005}. These studies have produced structures resembling the ones found in nature for different salts like NaCl, NaZn$_{13}$, AlB$_{2}$ \cite{Trizac1997, Bartlett1992, Vermolen2009, Hunt2000, Filion2011}, the Laves structures MgCu$_{2}$, MgNi$_{2}$, MgZn$_{2}$\cite{Sanders1978, Filion2011, Filion2011b}, and most recently, stable ISS\cite{Filion2011,Filion2011b}. This large diversity of equilibrium structures highlights their potential for applications in photonics, optics, semiconductors and structure design \cite{Trizac1997,Vermolen2009,Filion2011,Hunt2000}. In addition, due to their simplicity, binary hard sphere mixtures are ideal models to study the kinetics of crystallisation
in salts, metal alloys, metallic glasses and any other crystallising system
where there is more than one species, and so there is a compositional variable \cite{Filion2011,Filion2011b,Zhang2014,Royall2015,gludovatz14}.

However, only a few experimental studies have focused on the crystallisation \emph{process} of binary hard spheres, which can be due to the difficulty of obtaining close-packed ordered structures. This is related to slow kinetics, caused by a competition of the growth of crystal nuclei and compositional fluctuations. In addition, the differences in sedimentation rates between the particles also play an important role in preventing crystallisation\cite{Royall2008,Vermolen2008, Velikov2002, Hunt2000,Eldridge1995}. 
Similar to one-component systems, the particles in the crystalline structure have a higher translational entropy than in the metastable fluid, which compensates for the decrease of entropy as the system becomes more ordered \cite{Trizac1997,Schofield2005,Hunt2000,Bartlett1992,Eldridge1995}. In addition, it has been proposed that the theoretically predicted structures will only be thermodynamically stable if their maximum close packing fraction exceeds the correspondent 0.7405 for fcc or hcp lattices of one-component systems \cite{Schofield2005, Hunt2000,Trizac1997,Eldridge1995}. \par
Interestingly, these studies have also shown some discrepancies between the experimental observations and the predicted assemblies for particular size ratios and compositions, especially at concentrations near the glass transition \cite{Schofield2005,Filion2011,Eldridge1995}. Furthermore, the kinetic contributions to crystallisation, along with microscopic mechanisms that yield the ordering of mixtures are not yet fully understood \cite{Eldridge1995}, accentuating the complexity of these systems over one-component ones, and limiting their promising applications. 
Based on the interest in understanding such crystallisation processes, our goal is to study experimentally the heterogeneous crystallisation of a mixture of colloidal hard spheres at the particle-resolved level and compare our results with previous predictions. We also use simulations to study the evolution of the crystals, in particular the dynamics of the small particles, and propose a mechanism for the crystal formation. We observed that only interstitial solid solutions (ISS) are formed regardless of the different packing fractions tested in both experiments and simulations, as predicted \cite{Filion2011b}. In such structures, the crystalline phase --a mixture of fcc and hcp lattices-- is formed only by the large particles, whereas the small particles are positioned in
a fraction of the octahedral holes of the ordered phase. We compared the composition of our ISS with the equilibrium composition previously predicted for a related system\cite{Filion2011b}, and we observed interesting discrepancies at high packing fractions. By studying the dynamics of the system we found a trapping effect of the small particles which could be preventing further crystallisation and equilibrium filling of the octahedral holes. 

This paper is organised as follows. In section \ref{Methods} we describe the experimental and simulations details, along with the techniques used for the analysis of the ordered phases. In section \ref{ResandDis} we first present the results for crystallisation for our experiments and their structural analysis. We compare such results with previous work. We continue with the structural analysis of the ordered phases found on the simulations and the study of the evolution and quality of the crystals. We then focus on the dynamics of the species. We finalise by proposing a crystallisation mechanism for our system following the results obtained through simulations. Finally, in section \ref{Concl} we present our conclusions. 

\begin{figure*}[!t]
\includegraphics[width=1\textwidth]{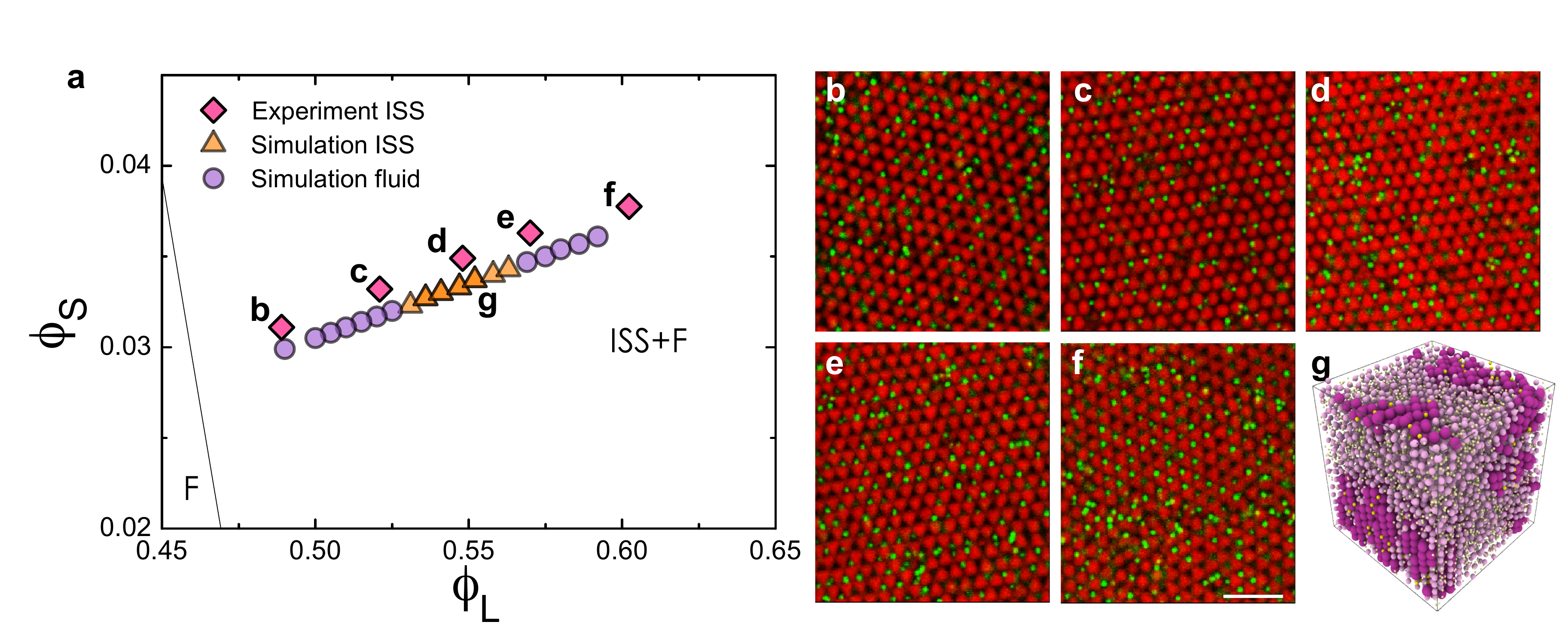}
\centering
\caption{\footnotesize{(a) The state points we study (symbols) at a size ratio $\gamma=0.39$, and number ratio of one, and Filion, \emph{et al.} phase diagram of binary hard spheres with a size ratio $\gamma$ = 0.4\cite{Filion2011b}, in the $\phi_\mathrm{{L}}$-$\phi_\mathrm{{S}}$ plane.
Diamonds: ISS from heterogenous nucleation experiments. Purple circles: simulations
that 
did not crystallise. 
Orange triangles: ISS from simulations.  (b-f) are snapshots of crystalline structures found on experiments for different $\phi_\mathrm{{tot}}$, the letters b to f in (a)
indicates the position on the phase-diagram.
The red and green colours are the large and small particles,
respectively, and the scale bar = 10 $\mu$m.
(g) is an ISS found for simulations with $\phi_\mathrm{{tot}}$=0.586, where crystalline large and small particles are coloured purple and yellow, respectively, whereas fluid large and small particles appear in light pink and light yellow, respectively.}}
\label{fig:PhaseDiagram-exp_sim}
\end{figure*}

\section{Methods} \label{Methods}

\subsection{Experiments}
The binary system used in the experiments consisted on sterically stabilised poly(methylmethacrylate) (PMMA) particles fluorescently labeled with rhodamine and coumarin dyes to enable separate fluorescent imaging. The sizes of the big particles have a diameter of 1700 nm with a polydispersity of 7\%, whereas the small particles are 669 nm in diameter and 7.8\% polydispersity, obtained by scanning electron microscopy. In spite of their large polydispersity, both particles are able to crystallise on their own. These dimensions yield a size ratio ($\gamma=\sigma_\mathrm{{S}}/\sigma_\mathrm{{L}}$) of 0.39. The dried particles were suspended separately in a solvent mixture of 27\% w/w cis-decahydrophnaphtalene (cis-decalin) and cyclohexyl bromide (CHB) that matches the density and refractive index of the particles. Additionally, tetrabutylammonium bromide (TBAB) salt was added in order to screen the charges, conditions that allow the particles to behave in a very close manner to hard spheres \cite{Royall2013,Vermolen2009, Royall2007}.
The suspensions were left to equilibrate for several hours, after which they were mixed together at a fixed number particle ratio, $N_{L}/N_{S}$=1, and at several total volume fractions, $\phi_\mathrm{{tot}}$, defined as $\phi_\mathrm{{tot}}=\phi_\mathrm{{L}}+\phi_\mathrm{{S}}$, where $\phi_\mathrm{{L}}$ and $\phi_\mathrm{{L}}$ refer to the partial packing fractions of each species. These were calculated based on the amount of particles and solvent weighed, following $\phi_\mathrm{{L/S}}=m_\mathrm{{L/S}}/(m_\mathrm{{L}}+m_\mathrm{{S}}+m_\mathrm{{solvent}})$. After shaking, the samples were confined to squared glass capillaries of 0.50 x 0.50 mm (Vitrocom) and sealed at each end with epoxy. The samples were studied by means of confocal laser scanning microscopy, CLSM (Leica SP5 fitted with a resonant scanner) using two different channels at 543 nm and 488 nm and NA 63x oil immersion objective. For particle tracking, 3D data sets were recorded by taking a full scan of the capillary in the $z$ axis, making sure the pixel size was equal in all axes (0.125 $\mu$m/pixel). We used particle resolved studies to determine the crystalline structure \cite{Ivlev2012}. The crystallisation time was calculated in Brownian time units for the large particles, i.e., the time it takes for the large spheres to diffuse a particle radius, given by $\tau_{B}=(\sigma_\mathrm{{L}}/2)^2/6D = 0.963$ s, where \textit{D} is the diffusion constant. The Brownian time $\tau_B$ thus sets our unit of time. Observations were carried out for over a month, which corresponds to $2.6\times10{^6}$ $\tau_{B}$, and were made all along the length and height of the capillaries, which were left standing in a vertical position.

\subsection{Simulations} 
Hard-sphere simulations were carried out using the open-source event-driven molecular dynamics DynamO package\cite{bannerman2011,Bannerman2009} in isothermal-isochoric (NVT) conditions for a binary mixture. The total number of particles is {$\mathrm{N_{tot}}$}= 10 968.
Particles of both species are of equal mass, $m=1$. The size and number ratio, as well as the final total particle densities were chosen to match the ones from the experiments, thus $\gamma$= 0.39 and $\phi_\mathrm{{tot}}$ from 0.52 to 0.64, respectively. A fluid system at $\phi_\mathrm{{tot}}$  = 0.282 was used as the initial configuration, which was then linearly compressed following
Stillinger and Lubachevsky\cite{Stillinger1993} to the appropriate 
packing fractions until an equilibrated structure was found. Experiments of colloidal systems have shown that the particles stick irregularly on the walls of the cell they are contained, thus producing an irregular surface over which the crystallisation takes place\cite{Royall2003}. Due to limitations in reproducing said phenomenon and thus the final surface, we elected to conduct our simulations with periodic boundary conditions. Future work could focus on simulating such environment and its influence over the kinetics of the system presented herein. 

Finally, simulation times were scaled to the experimental data. We inferred our $\tau_B$ in the simulations through the relationship $\tau_{\alpha}$= 2.597 $\tau_{B}$, obtained previously by comparing event-driven MD simulations and colloidal experiments of a one-component hard sphere system at $\phi$=0.38, where the structural relaxation time $\tau_{\alpha}$ of the simulations was found to equal 0.404 simulation time units\cite{Royall2014}. Since this mapping pertains to a one-component system, here we confirm that the relaxation of the large particles is similar in the binary system of interest herein. Therefore, we calculated $\tau_{\alpha}$ of the large particles from the trajectories at the same packing fraction ($\phi_\mathrm{{tot}}$=0.38). This we did by computing the intermediate scattering function (ISF), $F(\mathrm{k},t) = N^{-1}\big<\sum_{j=1}^{N}\mathrm{exp}[-i\mathrm{k}\cdot(x_{j}(t)-x_{j}(0)\big>$. To characterise the mobility in the length scale of a particle diameter, we evaluated $F(k,t)$ at the wavenumber $k= 2\pi/\sigma_\mathrm{{L}}$\cite{Royall2014,Dunleavy2015}. We obtained a value of $\tau_{\alpha}$=0.526 simulation time units, which is comparable to the one obtained by the mentioned work, thus confirming the validity of our approach, and enabling us to use the aforementioned relationship to infer the $\tau_B$ for our simulations. 

\subsection{Location of particles in experiment}

In order to determine the local ordering of the crystals formed in our experiments, we analysed separately the structures formed by each component through particle tracking. To locate the particles in the experiments, 3D data sets were directly analysed using the algorithm described in Leocmach and Tanaka \cite{Leocmach2013}. 

\subsection{Identification of local crystal structure}

Analysis of the crystalline structure for both experiments and simulations, was done following Lechner and Dellago \cite{Lechner2009} by obtaining the bond orientational order parameters (BOO), based on complex spherical harmonics. This analysis gives information of 
the degree and type of ordering, thus enabling us to differentiate between distinct crystalline phases. We focused on the locally averaged order parameters $\bar{q}_4$ and $\bar{q}_6$ for square and hexagonal orders, respectively. These parameters take into account the effect of the second nearest neighbours, which distinguish more clearly amongst different arrangements\cite{Lechner2009}. The methodology followed is detailed elsewhere\cite{Lechner2009,Jungblut2011,RiosdeAnda2015}. Briefly for each particle
$i$ and regardless of its small or large nature, we identify the nearest neighbours $N_b (i)$ using a cutoff corresponding to the contact distance between the closest large particles (obtained through the pair distribution function, $g(r)$). For the experiments said distance corresponded to 1.38$\sigma_\mathrm{{L}}$, whereas for the simulations this number was 1.25$\sigma_\mathrm{{L}}$. Finally for the perfect (by construction) NaCl-type lattice the cutoff was 1.1$\sigma_\mathrm{{L}}$. Using said list of neighbours, the local order parameters, or Steinhardt order parameters can be obtained:
\begin{equation}
q_{lm}(i)=\frac{1}{N_{b}(i)}\sum_{j=1}^{N_{b}(i)}Y_{lm}(\bm{r}_{ij}) \;,
\end{equation}
and summing over all the harmonics:
\begin{equation}
{q}_{l}(i)=\sqrt{\frac{4\pi}{2l+1}\sum_{m=-l}^{l}|q_{lm}(i)|^{2}} \;
\end{equation}
where $Y_{lm}(\bm{r}_{ij})$ correspond to the spherical harmonics, $l$ and $m$ are integer parameters, the latter running from $m=-l$ to $m=+l$, and $r_{ij}$ corresponds to the vector from particle $i$ to particle $j$. However, these parameters only contain information about the first shell surrounding particle $i$. In order to obtain information about the second shell, and thus, the locally averaged order parameters, we need to sum over the list of $\tilde{N}_b(i)$ of particle $i$ and the particle $i$ itself, giving: 
\begin{equation}
\bar{q}_{lm}(i)=\frac{1}{\tilde{N}_{b}(i)}\sum_{k=0}^{\tilde{N}_{b}(i)}q_{lm}(k),
\end{equation}
and again summing over all the harmonics we get:
\begin{equation}
\bar{q}_{l}(i)=\sqrt{\frac{4\pi}{2l+1}\sum_{m=-l}^{l}|\bar{q}_{lm}(i)|^{2}} \;
\end{equation} 

\subsection{Determination of the Lengthscale of Crystalline Ordering}

Following the analysis described in \cite{Fullerton2016,Klotsa2013} for identifying the size of domains, we determined in both experiments and simulations the typical crystalline domain size as $N\textsubscript{Q}=N^{-1}\big<\vec{Q}^{*}\cdot\vec{Q}\big>$, where $N$ is the total number of particles.
Here
$\vec{Q}=\sum_{p=1}^{N} \vec{q}_6(p)$ is the sum over all particles of the normalised complex vector $\vec{q}_6(p)$, whose components are the spherical harmonics with $\it{l}$=6. $\vec{q}_6(p)$ normalisation is done by setting
the average of
$\vec{q}_6(p)\cdot\vec{q}_6(p)^{*}=1$. The values taken by $N\textsubscript{Q}$ will depend on the amount of particles oriented in the same fashion.
The value of N\textsubscript{Q} is an estimate of the average
number of particles in a coherently ordered crystallite.

\begin{figure}[h!]
\includegraphics[width=0.4\textwidth]{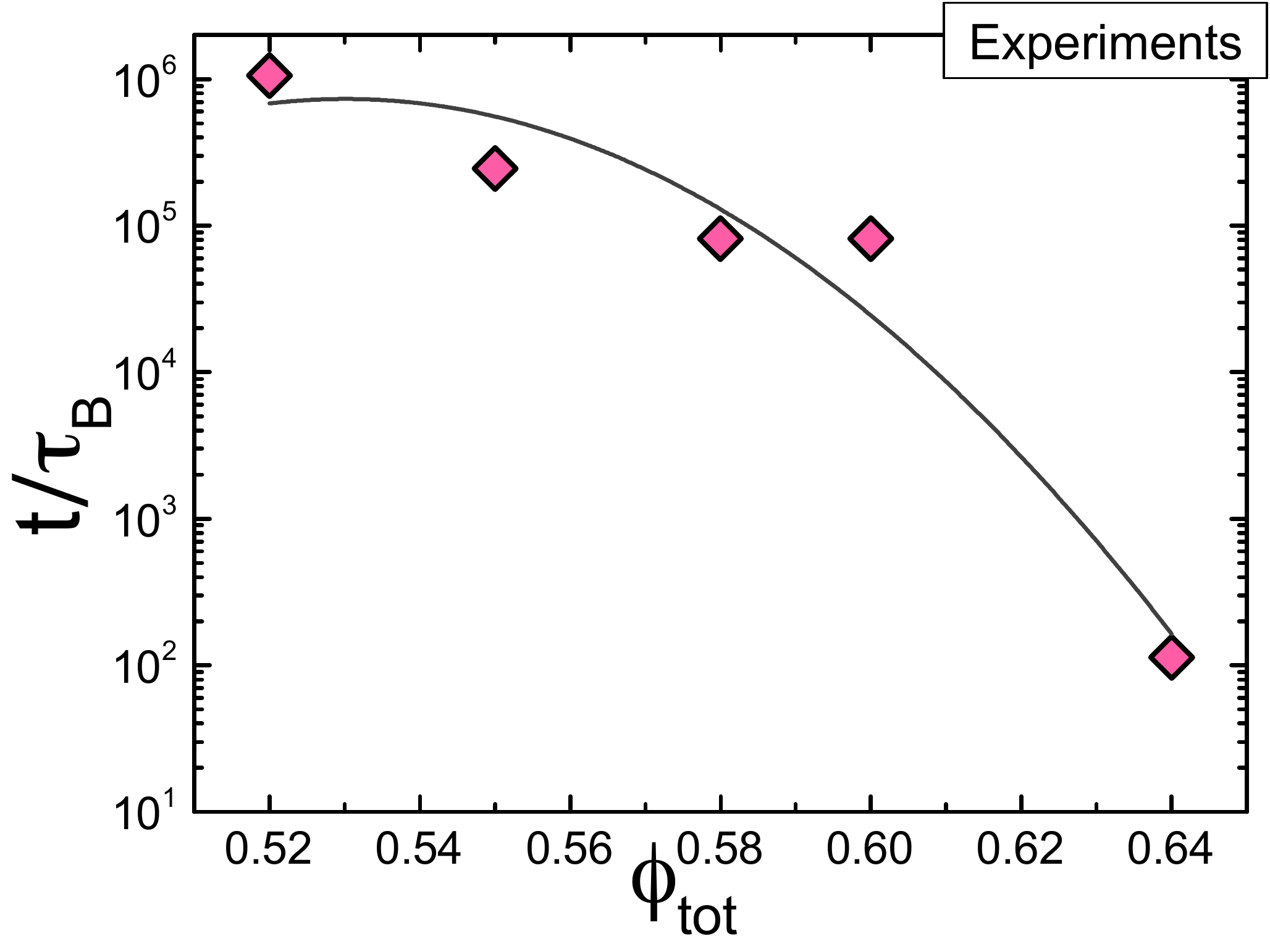}
\centering
\caption{\footnotesize{Experimental results of the heterogeneous crystallisation time of the binary system in terms of Brownian motion, $\tau_{B}$. The line is a guide to the eye.}}
\label{fig:Crystallisation_Time_Brownian}
\end{figure}

\section{Results and Discussion} \label{ResandDis}

We present our analysis in three sections: we first identify the ordering and quality of the structures found in both our experiments and simulations (Sec. \ref{IIIA}). 
We then focus on the dynamics of our two species in both the fluid and the crystalline regions in our simulations (Sec. \ref{IIIB}). Finally, we compare the composition of our ISS with an equilibrium system with a comparable size ratio and propose a mechanism for the formation of the structures found (Sec. \ref{IIIC}).

\begin{figure*}[!ht]
\includegraphics[width=0.7\textwidth]{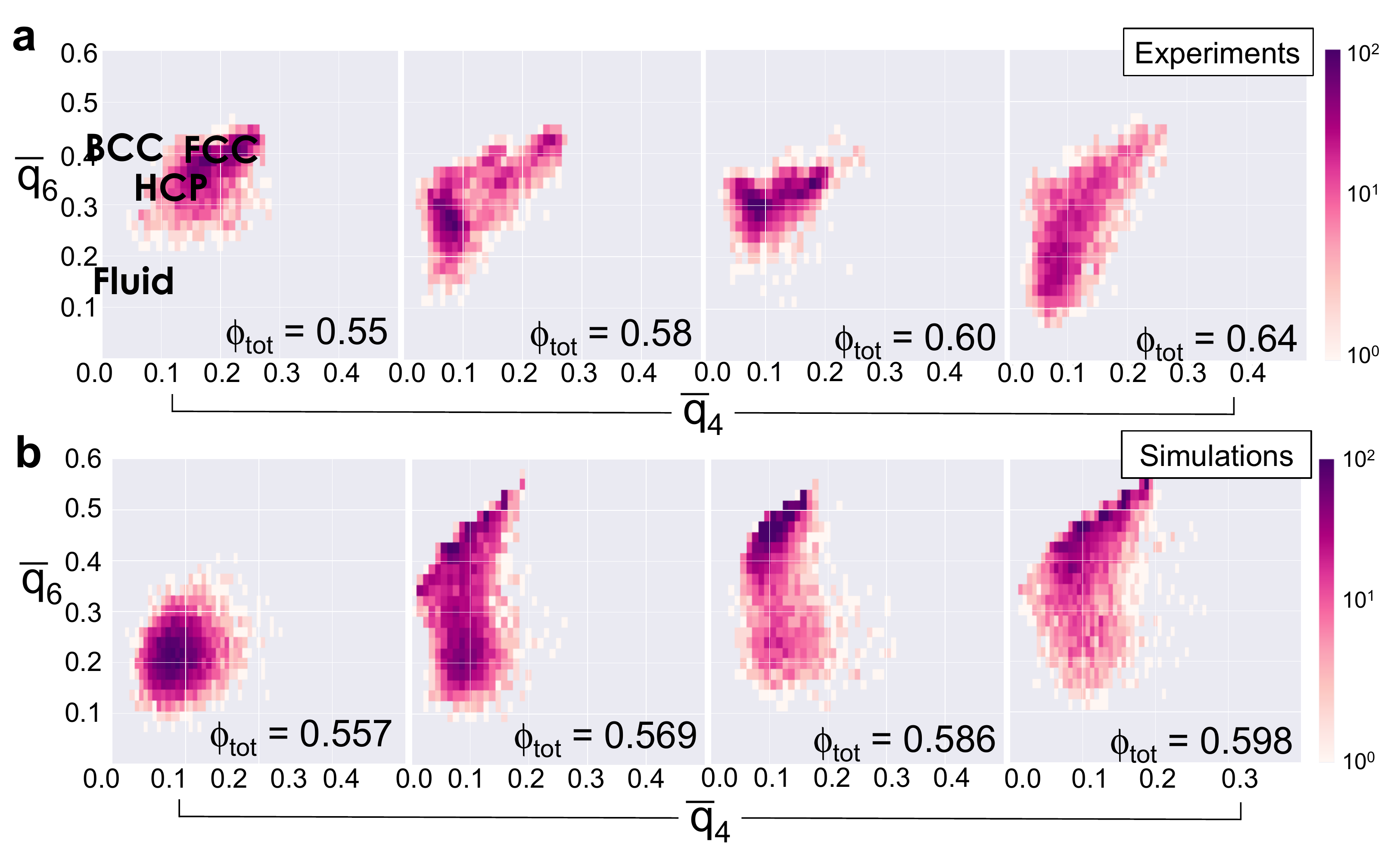}
\caption{\footnotesize{Local bond order parameter diagrams for the crystalline phase formed found in (a) experiments and (b) simulations, showing fcc, hcp and fluid coexistence for different total volume fractions tested for both studies, all at final times. $\phi_\mathrm{{tot}}$=0.557 (simulations) is included to show the presence of only a liquid phase. The regions for perfect bcc, fcc, hcp lattices are shown in the top-left plot. Both parameters were calculated using only the large particles. 
}}
\label{fig:Q6Q4-all}
\end{figure*}

\subsection{Analysis of the crystalline structures: identification of Interstitial Solid Solutions (ISS)} \label{IIIA}

Filion and collaborators used computer simulations to study a mixture of binary hard spheres with a  size ratio \cite{Filion2011b} only slightly higher (0.4) than ours (0.39). They calculated the equilibrium  phase diagram, which we reproduce in Fig. \ref{fig:PhaseDiagram-exp_sim} (a). There, the symbols represent our experimental state points for the experiments on heterogeneous nucleation (diamonds) and simulations (circles and triangles). Filion 
predicted that at equilibrium, there is coexistence between interstitial solid solutions (ISS), and a fluid phase (F)\cite{Filion2011b}. The ISS consists of a regular close-packed lattice, with the small particles filling some but not all the octahedral holes; the fraction of these holes varies continuously with composition.

Indeed we do observe crystalline structures built up by the large particles with interstitial small particles positioned in a random fashion (Snapshots Fig. \ref{fig:PhaseDiagram-exp_sim}), regardless of the $\phi_\mathrm{{tot}}$ and in coexistence with a fluid phase, for both the experiments and the simulations. Surprisingly, we did not find ordered structures for the simulations below $\phi_\mathrm{{tot}}$=0.563 or above $\phi_\mathrm{{tot}}$=0.610, where the mixtures did not crystallise (Fig. \ref{fig:PhaseDiagram-exp_sim} purple circles). 


In our experiments, nucleation of the crystals is heterogeneous,
the crystals grew parallel to the walls of the capillaries and the fluid phase was present far away from the walls. The presence of a flat wall clearly facilitates the nucleation process and thus the mixture is able to crystallise in a $\phi_\mathrm{{tot}}$ range larger than the simulations, where the nucleation starts in the bulk.

The ordered structures found are stable ISS\cite{DeCristofaro1977}, on the timescales we consider. In these, the crystals grow from the fluid phase and form ordered structures composed of the larger particles forming a close packed crystal, and the smaller particles in the interstices. The composition of these crystals is much poorer in the small particles than it is in the original fluid, where the number ratio is one. \cite{Campbell2008,Filion2011}


The heterogeneous crystallisation of our mixtures was studied as a function of the total volume fractions $\phi_\mathrm{{tot}}$ = $\phi_{L}$ + $\phi_{S}$ and we found that increasing this parameter increases the crystallisation rate, as shown in Fig. \ref{fig:Crystallisation_Time_Brownian}. This observation is in agreement with one component systems \cite{Sandomirski2011}.

\subsubsection{Local structure of the large particles}

For all of our samples, the equilibrium structure expected is an interstitial solid solution in coexistence with a fluid\cite{Filion2011b}. In such assemblies, the large particles are expected to form a mixture of hcp and fcc lattices, known as random hexagonal close packing (rhcp), since the free energy difference between these two crystalline phases is very small\cite{Filion2011,Filion2011b}.
The results of the analysis of the crystals formed only by the large particles for the experiments and simulations are summarised in Fig. \ref{fig:Q6Q4-all} (a and b, respectively), where we can observe that in both cases the crystalline phases do consist of a mixture of fcc and hcp lattices in coexistence with a fluid phase for all the total volume fractions tested. This random stacking has been identified before in experiments of heterogeneous nucleation of one-component systems \cite{Sandomirski2011}.



\begin{figure*}[!ht]
\includegraphics[width=0.9\textwidth]{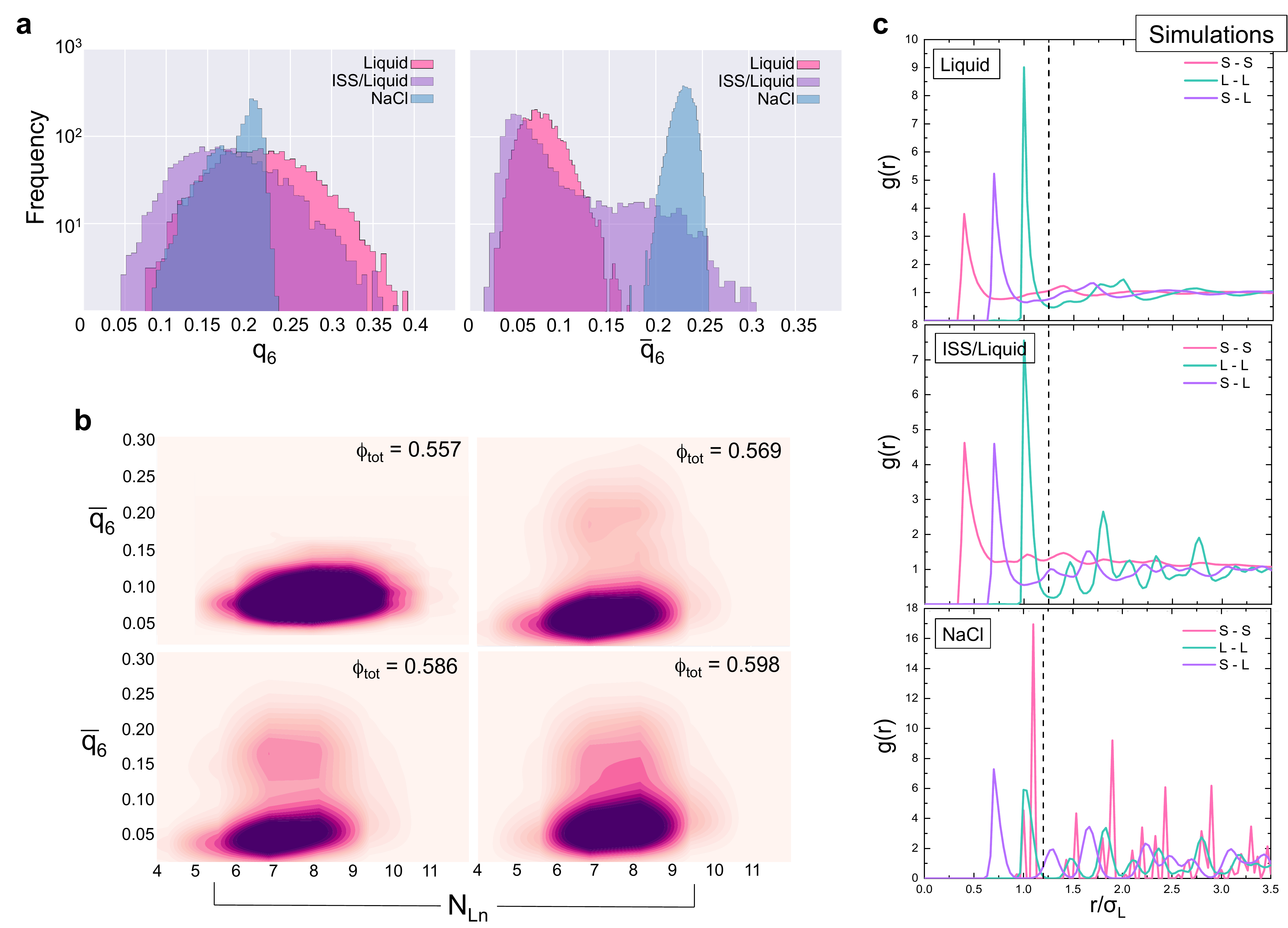}
\centering
\caption{\footnotesize{(a) Histograms of the bond orientational parameters $q_6$ (left) and  $\bar{q}_6$ (right) values of the small particles at $\phi_\mathrm{{tot}}$=0.586 before crystallisation (Liquid) and after crystallisation (ISS/Liquid), and an equilibrated NaCl-type crystal at $\phi_\mathrm{{tot}}$=0.64, showing the ability of $\bar{q}_6$ to distinguish the local environment of the small particles. (b) Heat maps of the $\bar{q}_6$ values of the small particles and their corresponding number of neighbouring large particles (N\textsubscript{Ln}) for different total volume fractions that crystallised in simulations, calculated following reference 33. The plots show two populations: one below and one above $\bar{q}_6$ = 0.15. Note that $\phi_\mathrm{{tot}}$=0.557 does not crystallise. (c) Radial distribution functions of the liquid (top) and ISS/Liquid (middle) both at $\phi_\mathrm{{tot}}$=0.586, and NaCl-type crystal (bottom) at $\phi_\mathrm{{tot}}$=0.64. Each shows the small-small (S-S), large-large (L-L) and small-large (S-L) interactions. The dashed lines indicate the minimum corresponding to the small and large contact cutoff used for the BOO analysis.
}}
\label{fig:q6vaq6_gr_NLn}
\end{figure*} 

\subsubsection{Local environment of the small particles: an analysis on the crystal quality and the vacancies in simulations}

In the ISS the small particles are situated in the octahedral holes of the fcc and hcp lattices formed by large particles. Filion \emph{et al.} showed, for
their smaller size ratio of 0.3, that the small particles are able to hop at lower pressures-- where the system is far from close-packing-- fill all the octahedral holes and yield an ISS
with the NaCl structure\cite{Filion2011}. On the other hand, for the larger size ratio of 0.4, Filion and coworkers also found that the ISS are the stable assembly, however, for this larger size ratio they do not report the formation of a perfect crystalline NaCl structure, nor they show the dynamics of the small particles like for $\gamma$=0.3\cite{Filion2011b}. In our experiments a similar size ratio of 0.39, we also observed only ISS, and we could not observe hopping of the small particles in direct imaging. Due to the limitations on tracking the motion of the small particles, we decided to conduct hard sphere molecular dynamics simulations in order to quantify both the small sphere
ordering and dynamics.

Using simulation data, we calculated the BOO $q_6$ and $\bar{q}_6$ values for the small
component, see Fig.~\ref{fig:q6vaq6_gr_NLn}(a) and (b). 
In Fig. \ref{fig:q6vaq6_gr_NLn}(a), we plot $q_6$ (left) and $\bar{q}_6$ (right) in the
liquid state, in a state that is partially crystallised with remaining liquid, and in a
perfect NaCl-type crystal. It is clear that $q_6$ does not differentiate between the different configurations, but $\bar{q}_6$ does distinguish between the liquid and the ISS. So we
use the Lechner-Dellago averaged $\bar{q}_6$ in order to identify the fluid or crystalline nature of the particles. Values of $\bar{q}_6$ above 0.15 are very rare
in the liquid, while in a well-ordered NaCl structure $\bar{q}_6$ is almost always above 0.15.

In Fig.~\ref{fig:q6vaq6_gr_NLn}(b), we show heat maps of the probability distribution
$p(N_{\mathrm{Ln}},\bar{q}_6)$ for 4 states, one liquid (top-left) and three ISS/liquid
mixtures. The samples with liquid and ISS show two populations. The dominant one, with $\bar{q}_6< 0.15$ is the liquid, but there is a smaller population with $\bar{q}_6>0.15$, which is consistent with values of $\bar{q}_6$ in the crystal. The association between this second population and the crystal is further supported by the fact that small particles here have around 6 to 8
large-particle neighbours. In the NaCl structure, each small particle has 6 large-particle nearest neighours, and our cutoff distance for neighbours (dotted line in Fig.~\ref{fig:q6vaq6_gr_NLn}(c))
is above the small-large nearest neighbour distance, so we would find at least 6 neighbours --- precisely what we observed.

\begin{figure*}[t!]
\includegraphics[width=1\textwidth]{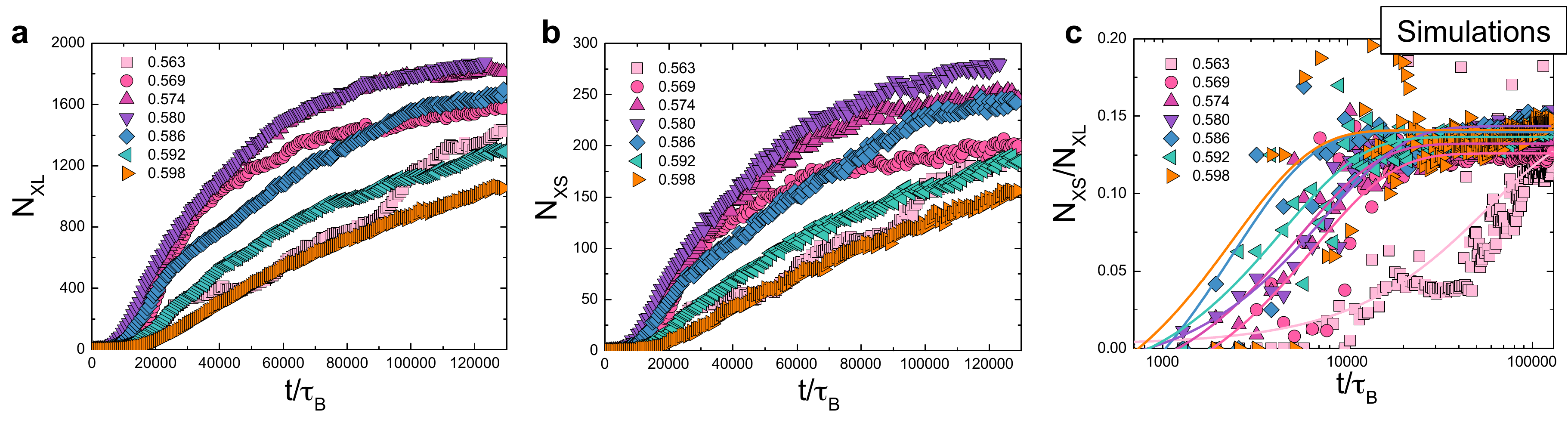}
\centering
\caption{\footnotesize{The numbers of large (a) and small (b) crystalline particles as a function
of time in simulations. (c) evolution of the proportion of small crystalline particles relative to the large crystalline ones in simulations according to their $\phi_\mathrm{{tot}}$.The simulation time has been rescaled to Brownian time 
(see Methods for details).}}
\label{fig:Simulation-Summary-NX}
\end{figure*}

\subsubsection{Composition of the crystalline domains}

We followed the crystal growth by estimating N\textsubscript{XL} and the small particles located at the octahedral holes (N\textsubscript{XS}), according to their $\bar{q}_6$ values. These results are shown in Fig. \ref{fig:Simulation-Summary-NX} (a) and (b), respectively. Here, we observe that both quantities reach a plateau, indicating no further crystal growth and suggesting no hopping of the small particles within the interstitial sites of the fcc and hcp lattices, as discussed previously\cite{Filion2011}. Next, we calculated the amount of N\textsubscript{XS} relative to N\textsubscript{XL} forming the ordered structure. We found two notable 
phenomena. Firstly, the octahedral holes are filled rapidly as the nucleation takes place (Fig. \ref{fig:Simulation-Summary-NX} (c)) with a maximum filling of the octahedral sites of around 14\%, above $\phi_\mathrm{{tot}}$=0.586 (see Fig. \ref{fig:Composition} (b)). Secondly, we observed no further changes in the proportion of crystalline particles, suggesting a trapping effect. Since the number of small crystalline particles also reached a constant value, we infer that they become trapped (and immobilised) in the the crystalline structure once it is formed. These phenomena will be investigated in detail in the following sections.

\subsubsection{Crystalline domains in experiment and simulation} \label{Ordering}

\begin{figure}[!hb]
\includegraphics[width=0.43\textwidth]{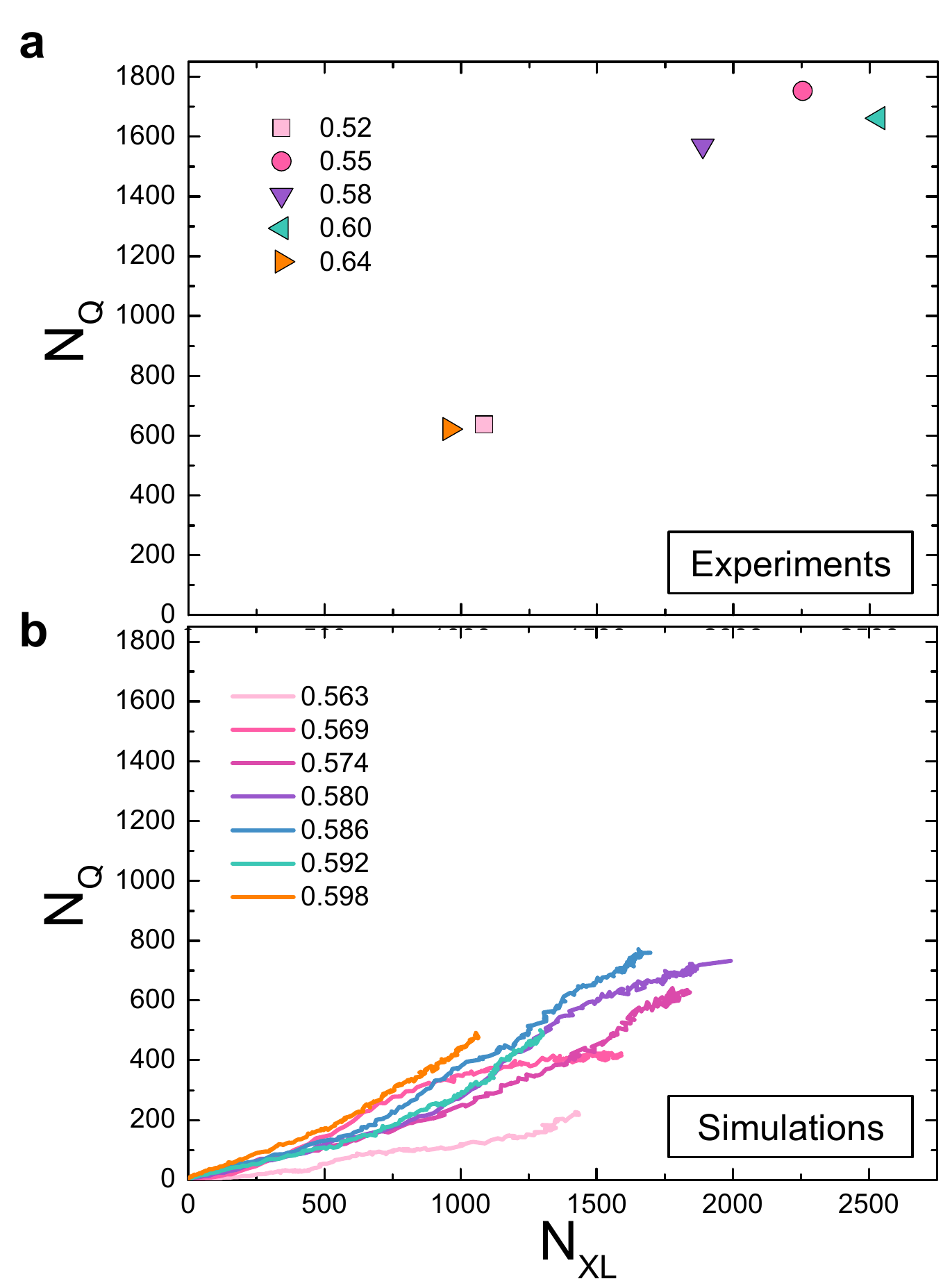}
\centering
\caption{\footnotesize{
The characteristic size of the crystal domain(s), N\textsubscript{Q},
as a function of the total number of large crystalline particles, N\textsubscript{XL}.
For the experiments, we show the final values (as symbols, see key), while
for computer simulations, we show the evolution with time during crystallisation
(as curves).
The total number of large particles in the experiments is 2534, 2645, 2584, 2524 and 2244 for $\phi_\mathrm{{tot}}$=0.64, 0.60, 0.58, 0.55 and 0.52, respectively. On the other hand, for simulations the total number is 5,484 of large and 5,484 of small particles. 
 N\textsubscript{Q} shows that the quality of the crystals obtained in the experiments is higher than the simulations. 
}}
\label{fig:XllinityvNq-both}
\end{figure}

Notably, see Figs. \ref{fig:PhaseDiagram-exp_sim} and \ref{fig:XllinityvNq-both}, the range of packing fractions that crystallised \emph{on the simulation timescale} was smaller than that of the experiments. But in both experiments and simulations, there are ranges
of volume fraction where maximal crystallisation occurs. In experiments
we observe regions in the system where there is largely complete crystallisation
of the large particles: at $\phi_\mathrm{{tot}}=0.55$, $0.58$ and $0.60$ ---see Fig.~\ref{fig:XllinityvNq-both} (a), the fraction of large particles identified
as crystalline is around 65\% or above.

In our simulations, the total amount of crystalline structure  is much smaller,
reaching the highest amount around 18\% at $\phi_\mathrm{{tot}}$=0.580, after which it started to decrease, as shown in Fig. \ref{fig:XllinityvNq-both} (b). 
The need for a higher $\phi_\mathrm{{tot}}$ to observe crystallisation on the simulation timescale ($\phi_\mathrm{{tot}}$=0.563 in comparison with $\phi$=0.53 for one component systems\cite{schilling2010}) might be due to a higher free energy barrier to generate nuclei necessary for crystallisation.
The drop of the crystalline fraction of our system at $\phi_\mathrm{{tot}}=0.598$
is similar to the one where one-component systems exhibit slow dynamics $\phi\simeq0.58$\cite{Royall2015}, and is related to the particles moving more slowly as their concentration increases, requiring a long time to rearrange in ordered structures. A similar trapping phenomenon could be happening here, where the particles do not move fast enough to rearrange and form crystals.

To determine if these crystalline large particles are in one large crystalline domain,
or if the system is polycrystalline, we follow 
Klotsa and Jack \cite{Fullerton2016,Klotsa2013}.
These authors defined a parameter N\textsubscript{Q} to describe the
size of coherent crystallites, i.e., the typical number of particles in a coherently ordered crystal.
We applied this analysis to both our experiments and simulations to determine which conditions yielded the crystals with the highest quality. In the case of the experiments, we analysed the final crystals formed by the large particles, whereas for the simulations, we studied the evolution of N\textsubscript{Q} through the crystallisation of the different samples.

The results are presented in Fig. \ref{fig:XllinityvNq-both}, where we compare this parameter with the number of large crystalline particles for both cases, N\textsubscript{XL}.
It is evident from these results that the size of the crystalline domains
obtained from heterogeneous nucleation in the experiments is higher than the homogenous nucleation present in the simulations, as N\textsubscript{Q} presents significantly larger values in the former for similar N\textsubscript{XL}. Furthermore, the N\textsubscript{Q} values for the experiments are close to the number of crystalline particles. 
Reasons for this include the presence of the flat walls of the capillaries, larger system sizes or the need for longer waiting times. Indeed flat walls can serve as a template able to enhance the nucleation and layering of the particles and thus improve the orientation of the crystal\cite{Sandomirski2011,Dullens2004}. On the other hand, estimations of the crystallisation time for our experiments for homogenous nucleation were found to be two orders of magnitude larger than the corresponding crystallisation time on our simulations, suggesting that longer waiting times are required for experimental homogenous crystallisation. This might be related to the higher polydispersity in the experimental system. Moreover, said time surpasses our limit for experimental timescale observations of a month ($2.6 \times 10^6$ $\tau_B$). \par
Finally, the system size in the experiments is of order $10^5$ times bigger than the simulations. Hence it seems likely that our experiments can crystallise over a wider range and form crystals with a greater quality due to a combination of the reasons mentioned above. 

In the case of simulations, the numerical values of N\textsubscript{Q} are significantly smaller than N\textsubscript{XL}, suggesting the presence of several clusters in all the samples. Also, a non-monotonic behaviour of this value is noted, since the highest values of the former do not correspond to the largest ones for the latter. This is evident for $\phi_\mathrm{{tot}}$=0.598, whose maximum N\textsubscript{XL} surpasses the corespondent of $\phi_\mathrm{{tot}}$=0.563 and 
0.569, with higher amounts of crystalline particles (Fig. \ref{fig:XllinityvNq-both} (b) orange, light pink, and pink lines, respectively). Interestingly, the dependence between these two parameters is not linear, with N\textsubscript{Q} increasing faster than the crystal growth. A possible reason for this could be fast coalescence between the forming crystallites. Indeed, $\phi_\mathrm{{tot}}$=0.574 (Fig. \ref{fig:XllinityvNq-both} (b) dark pink line) shows a faster increase of N\textsubscript{Q} around N\textsubscript{XL}$\sim1000$, which could be a consequence of two clusters merging to form a larger one.


\begin{figure}[ht!]
\includegraphics[width=0.43\textwidth]{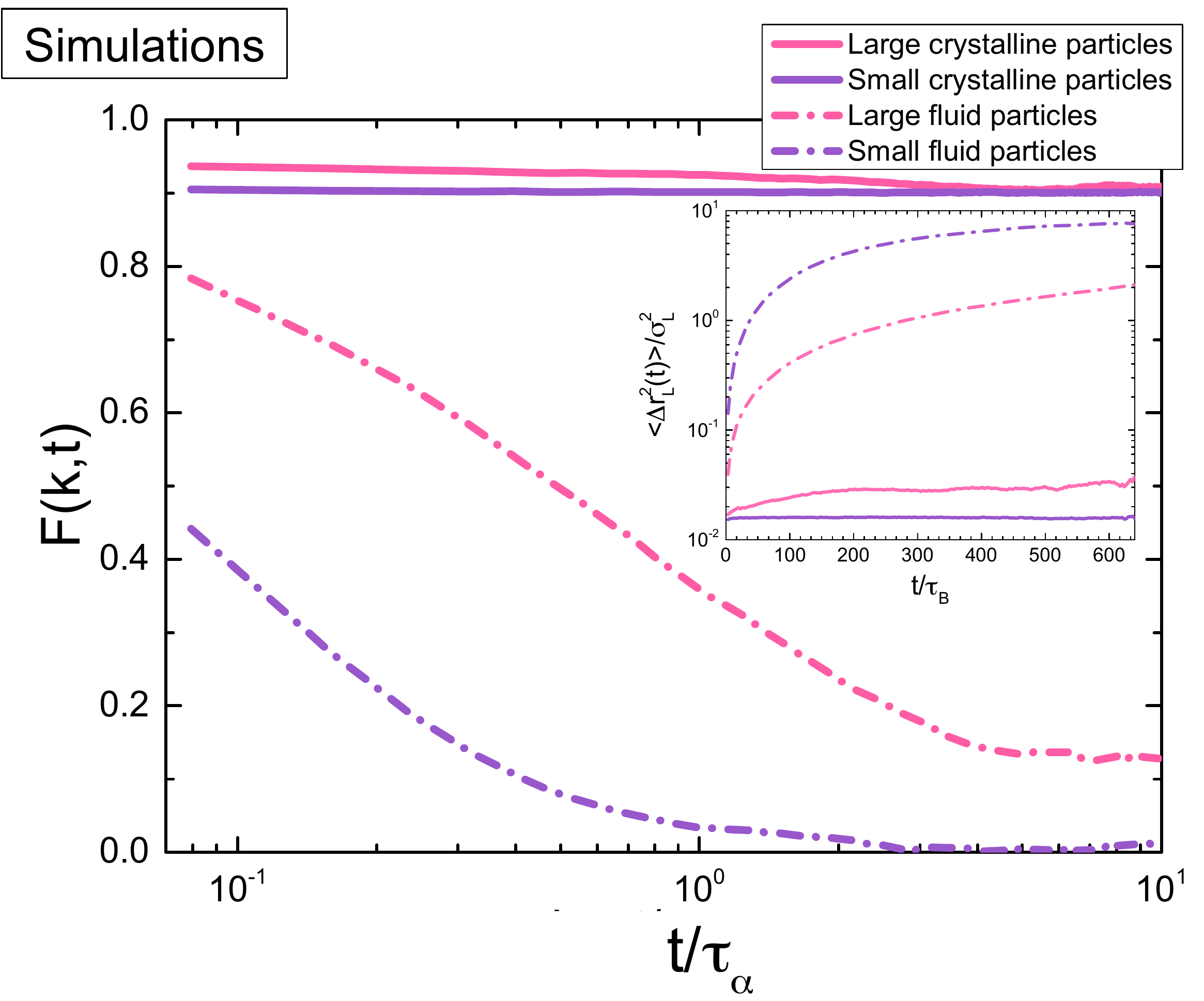}
\centering
\caption{\footnotesize{Intermediate scattering functions and mean square displacements (insert) from simulations, at $\phi_\mathrm{{tot}}$=0.586.
The large and small particles are shown in pink and purple, with continuous lines for
particles identified as crystal, and dotted lines for particles identified as fluid. The simulations start at the end of a crystallisation simulation. Particles at the liquid-crystal boundary region were discarded in order to characterise only the crystalline and fluid particles.}}
\label{fig:XtalvsFluid}
\end{figure}

\subsection{The dynamics of both large and small particles
are arrested in the crystals} \label{IIIB}

In order to show that our small particles were indeed localised, we computed both the mean square displacement (MSD, $\big\langle\Delta r^{2}_\mathrm{L}(t)\big\rangle = \big\langle (r_\mathrm{L}(t) - r_\mathrm{L}(0))^{2}\big\rangle$) and the intermediate scattering function (ISF), as described previously\cite{Dunleavy2015} for the simulations with $\phi_\mathrm{{tot}}$=0.586 for geometrically selected large and small particles in both liquid and crystalline regions. This was done at the final points of the simulation, where the ratio N\textsubscript{XS}/N\textsubscript{XL} remained constant. These results are shown in Fig. \ref{fig:XtalvsFluid}. We observe that both the large and small particles in the liquid region present movement (Fig. \ref{fig:XtalvsFluid}, dotted pink and purple lines, respectively), \emph{i.e.} their MSD (insert) and ISF present a behaviour typical for a confined random walk and fluids, respectively. In contrast, both species located in crystalline regions do not show any movement, identified as a flat line close to zero in the MSD and as a non-decaying line on the ISF (Fig. \ref{fig:XtalvsFluid}, continuous lines). We can conclude therefore that indeed, once the crystals are formed, the small particles are immobile within the crystal formed by the large particles; unable to hop and fill all the available octahedral holes, and thus our ISS remain as long-lived out-of-equilibrium structures.

\subsection{Comparison of our results at a size ratio \textbf{$\gamma$}=0.39
to those of Filion {\emph et al.} at size ratios of 0.3 and 0.4} \label{IIIC}

\begin{figure}[ht!]
\includegraphics[width=0.43\textwidth]{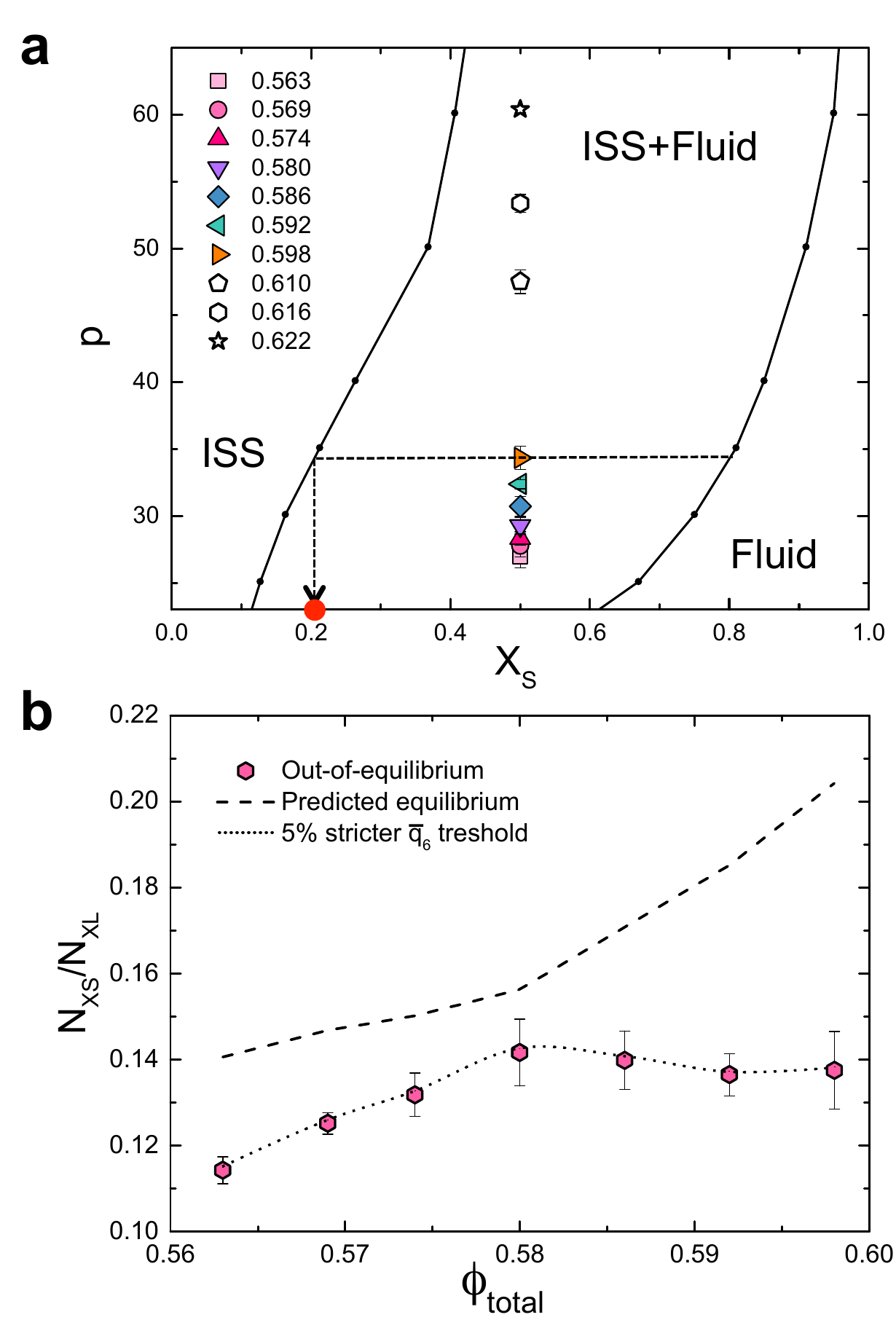}
\centering
\caption{\footnotesize{(a) Filion's\cite{Filion2011b} phase diagram of a binary hard-sphere system with $\gamma=0.4$, showing the reduced 
 coexistence pressure $p=\beta P \sigma^{3}$, as a function of composition $X_\mathrm{{S}}=N_\mathrm{{S}}/(N_\mathrm{{S}}+N_\mathrm{{L}}$. We have added to this filled symbols
to represent our state points where we obtained an ISS and the empty symbols for
the conditions were, on our time scales, we did not observe crystallisation. The tie line and arrow are drawn to illustrate the interpolation process to obtain the small particle composition in the equilibrium ISS. (b) The composition as a function of our system's average volume
fraction, $\phi_\mathrm{{tot}}$. The dashed line corresponds to the equilibrium
composition \cite{Filion2011b}, and the points are our simulation results. The error
bars are the standard deviation of 8 runs, whereas the dotted line is the error obtained when using a 5\% more strict  $\bar{q}_6$ threshold.}}
\label{fig:Composition}
\end{figure}


We found
that the small particles occupy the octahedral sites of the crystalline lattices in an incomplete fashion,
forming
a dilute ISS which in our computer simulations
consistently has a composition close to seven large particles
for every one small particle, much less than the one-to-one ratio
in the original fluid phase.

An ISS has also been identified before in binary mixtures 
with a smaller size ratio of 0.3 compared to 0.39 here
 \cite{Filion2011}.
  Filion and coauthors found that in order for the small particles to hop from one octahedral hole to the other, they needed to go through the adjacent tetrahedral hole
   ($\sim$0.225$\sigma_\mathrm{{L}}$ across).
 Although slow, they found that their smaller particles
could hop, whereas (see Fig. \ref{fig:XtalvsFluid}) our larger small particles
cannot. 

Filion also studied the size ratio 0.4, closer to the one used here. For that size ratio, the ISS equilibrium region in the phase diagram is larger than for 0.3. In order to confirm that our ISS are out-of-equilibrium, we compared our
simulation values of X\textsubscript{S}, with the equilibrium values determined
by Filion \cite{Filion2011b}. As shown in Fig. \ref{fig:Composition}(a), we determined
the final pressure in our simulations, and then using tie lines (which are horizontal
here as the two phases have the same pressure) read off the compositions of both phases.
 The comparison of these results and the ones corresponding to our ISS are shown in Fig. \ref{fig:Composition}(b), where the dashed curve is the
equilibrium X\textsubscript{S}, and the dotted curve and points are our results.
In all cases, the amount of small particles within the crystalline structure of large ones is smaller than the one predicted for an ISS in equilibrium. This difference is large
for systems at higher volume fractions.  

Thus we conclude that our ISS are out of equilibrium, and presumably are out of equilibrium
when they nucleate and grow.
As the large particles crystallise into an ordered structure, some of the small particles become trapped in the octahedral holes, forming the ISS.
But under the conditions we study, the integration of the small particles into
the growing crystal appears to be inefficient, resulting in the low number
of small particles in the crystal. This then causes the composition of the
remaining fluid to become increasingly rich in the small particles, as the crystal
grows. 

As we can see in Fig. \ref{fig:Composition}(a), increasing
X\textsubscript{S} will ultimately move the fluid out of the coexistence region and into the one-phase
fluid region of the phase diagram. So, we believe this change in composition
contributes to the growth of the crystal slowing
and stopping.
 
\section{Conclusions} \label{Concl}
Interstitial solid solutions (ISS) were identified in both particle resolved experiments and simulations of a binary mixture with a size ratio of 0.39. 
Through particle resolved studies carried out 
on the experimental results, we were able to identify that the crystalline structure was made up by the large particles forming a mixture of fcc and hcp lattices, with low crystallinity, and where the small particles are localised randomly within the octahedral holes. Simulations showed the same crystalline structure and allowed us to follow the crystallisation process and quantify the amount of filling by the small particles and, significantly, the dynamics of the small particles in the ISS. With this information, we were able to propose a crystallisation mechanism where the large particles form ordered structures independent containing small particles, which become rapidly trapped within the growing crystal to a maximum of $\sim14\%$, a smaller filling from the predicted in equilibrium conditions, particularly for packing fractions above 0.586. This exclusion of the small particles
depletes the fluid around the growing crystal and we believe
that this compositional change prevents further crystalline growth. Finally, we were able to show that the small particles are not able to penetrate the formed crystalline structure nor to move within the available octahedral holes, thus producing long-lived out-of-equilibrium ISS.  

\section{Acknowledgements}
The authors are grateful to John Russo for simulation discussions. IRdA would like to thank Conacyt for financial support. CPR acknowledges the Royal Society for funding and Kyoto University SPIRITS fund. FT and CPR acknowledge the European Research Council (ERC consolidator grant NANOPRS, project number 617266). This work was carried out using the computational facilities of the Advanced Computing Research Centre, University of Bristol.

\section{References}
\bibliographystyle{ieeetr}
\bibliography{factory.bib}

\end{document}